\documentclass[pra,aps,preprint,amsmath,amssymb,showpacs]{revtex4}
\usepackage{graphicx}
\usepackage{epsfig}
\include{graphics}
\begin{document}

\title{Many-body interaction in fast soliton collisions}
\author{Avner Peleg$^{1}$, Quan M. Nguyen$^{2}$, and Paul Glenn$^{1}$}
\affiliation{$^{1}$ Department of Mathematics, State University of New York
at Buffalo, Buffalo, New York 14260, USA}

\affiliation{$^{2}$ Department of Mathematics, International University, 
Vietnam National University-HCMC, Thu Duc, Ho Chi Minh City, Vietnam}


\begin{abstract}
We study $n$-pulse interaction in fast collisions 
of $N$ solitons of the cubic nonlinear Schr\"odinger (NLS) equation
in the presence of generic weak nonlinear loss. 
We develop a reduced model that  yields the contribution of 
$n$-pulse interaction to the amplitude shift for collisions in 
the presence of weak $(2m+1)$-order loss, for any $n$ and $m$. 
We first employ the reduced model and numerical solution of  
the perturbed NLS equation to analyze soliton collisions 
in the presence of septic loss $(m=3)$. 
Our calculations show that three-pulse interaction gives the 
dominant contribution to the collision-induced amplitude shift 
already in a full-overlap four-soliton collision, 
and that the amplitude shift strongly depends on 
the initial soliton positions. We then extend these results 
for a generic weak nonlinear loss of the form $G(|\psi|^{2})\psi$, 
where $\psi$ is the physical field and $G$ is a    
Taylor polynomial of degree $m_{c}$. Considering $m_{c}=3$, 
as an example, we show that three-pulse interaction gives the 
dominant contribution to the amplitude shift in a six-soliton 
collision, despite the presence of low-order loss.
Our study quantitatively demonstrates that $n$-pulse interaction 
with high $n$ values plays a key role in fast collisions of 
NLS solitons in the presence of generic nonlinear loss. 
Moreover, the scalings of $n$-pulse interaction 
effects with $n$ and $m$ and the strong dependence on initial 
soliton positions lead to complex collision dynamics,  
which is very different from the one observed in 
fast NLS soliton collisions in the presence of cubic loss.     
\end{abstract}
\maketitle

\section{Introduction}
\label{Introduction}
The problem of predicting the dynamic evolution of $N$ physical 
interacting objects or quantities, commonly known as the $N$-body 
problem, is an important subject of research in science and engineering. 
The study of this problem plays a key role in many fields, 
including celestial mechanics \cite{Roy2005,Hagihara70}, 
nuclear physics, solid-state physics, and molecular 
physics \cite{Thouless72}. In many cases, the 
dynamics of the $N$ objects is governed by a force, which is a 
sum over two-body forces. This is the situation in celestial 
mechanics \cite{Roy2005,Hagihara70} and in other systems \cite{Thouless72}, 
and it has been discussed extensively in the literature. A different but 
equally interesting dynamic scenario emerges when the $N$-body 
dynamics is determined by a force involving $n$-body interaction 
with $n\ge 3$ \cite{Definition}. Indeed, $n$-body forces with $n\ge 3$ 
have been employed in a variety of problems including 
van der Waals interaction between atoms \cite{Teller43}, 
interaction between nucleons in atomic 
nuclei \cite{Fujita57,Nogami67,Witala98,Witala2002}, and in cold atomic 
gases in optical lattices \cite{Daley2009,Honer2010,Liang2012}. 
A fundamental question in these studies concerns 
the physical mechanisms responsible 
for the emergence of $n$-body interaction with a given $n$ value.     
A second important question revolves around the dependence of the 
interaction strength on $n$ and on the other physical parameters.  
In the current study we investigate a new class of $N$-body problems, 
in which $n$-body forces play a dominant role. More specifically, 
we study the role of $n$-body interaction in fast collisions 
between $N$ solitons of the cubic nonlinear Schr\"odinger (NLS) 
equation in the presence of generic weak nonlinear loss. 
In this case the solitons experience significant 
collision-induced amplitude shifts, and important questions 
arise regarding the role of $n$-pulse interaction in the 
process, and the dependence of the amplitude shift and the  
$n$-pulse interaction on the physical parameters.

The NLS equation is one of the most widely used nonlinear wave 
models in the physical sciences. It was successfully employed 
to describe a large variety of physical systems, including 
water waves \cite{Zakharov84,Newell85}, 
Bose-Einstein condensates \cite{Dalfovo99,BEC2008}, 
pulse propagation in optical waveguides \cite{Hasegawa95,Agrawal2001}, 
and nonlinear waves in plasma \cite{Asano69,Ichikawa72,Horton96}. 
The most ubiquitous solutions of the NLS equation are 
the fundamental solitons. The dynamics of 
fundamental solitons in these systems can be affected 
by loss, which is often nonlinear \cite{Malomed89}. 
Nonlinear loss arises in optical waveguides due to 
gain/loss saturation or multiphoton absorption \cite{Boyd2008}. 
It is also quite common in certain Bose-Einstein 
condensates \cite{Burt97,PK2007} and in many other systems  
that are described by the complex Ginzburg-Landau 
equation \cite{Kramer2002}. It is therefore important to study 
the impact of nonlinear loss on the propagation of 
fundamental NLS solitons.

The main effect of weak nonlinear loss on the propagation 
of a single NLS soliton is a continuous decrease 
in the soliton's energy. This single-pulse amplitude shift 
is qualitatively similar to the one due to linear loss, 
and can be calculated in a straightforward manner by 
employing the standard adiabatic perturbation theory. 
Nonlinear loss also strongly affects the collisions of 
NLS solitons, by causing an additional decrease of soliton 
amplitudes. The character of this collision-induced amplitude 
shift was recently studied for fast soliton collisions in the 
presence of cubic and quintic loss \cite{PNC2010,PC2012}.  
The results of these studies indicate that the amplitude 
dynamics in soliton collisions in the presence of generic 
nonlinear loss might be quite complicated due to $n$-pulse 
interaction effects. More specifically, in Ref. \cite{PNC2010} 
it was shown that the total collision-induced amplitude 
shift in a fast three-soliton collision in the presence 
of {\it cubic} loss is given by a sum over amplitude shifts due to 
two-pulse interaction, i.e., the contribution to the amplitude 
shift from three-pulse interaction is negligible. In contrast, 
In Ref. \cite{PC2012} it was found that three-pulse 
interaction enhances the amplitude shift in a fast 
three-soliton collision in the presence of {\it quintic} 
loss by a factor of 1.38.

The results of Ref. \cite{PC2012} suggest that $n$-pulse 
interaction with $n \ge 3$ might play an important 
role in fast NLS soliton collisions in the presence of 
generic or high-order nonlinear loss. Despite of this 
fact, a systematic analytic or numerical study of the role of 
$n$-pulse interaction in these collisions is still missing.  
In the current study we address this important problem. 
For this purpose, we first develop a general reduced model 
for amplitude dynamics, which allows us to calculate the contribution 
of $n$-pulse interaction to the amplitude shift for collisions in 
the presence of weak $(2m+1)$-order loss, for any $n$ and $m$. 
We then use the reduced model and numerical solution of  
the perturbed NLS equation to analyze soliton collisions 
in the presence of septic loss $(m=3)$. 
Our calculations show that three-pulse interaction gives the 
dominant contribution to the collision-induced amplitude shift 
already in a full-overlap four-soliton collision.  
Furthermore, we find that the amplitude shift strongly depends on 
the initial soliton positions with a pronounced maximum in the case 
of a full-overlap collision. We then generalize these results 
for generic weak nonlinear loss of the form $G(|\psi|^{2})\psi$,  
where $\psi$ is the physical field and $G$ is a    
Taylor polynomial of degree $m_{c}$. We consider $m_{c}=3$, as an example.  
That is, we take into account the effects of linear, cubic, quintic, and 
septic loss on the collision. We show that in this case three-pulse 
interaction gives the dominant contribution to the amplitude shift 
in a six-soliton collision, despite the presence of linear and cubic loss. 
Our study uncovers a new type of $n$-body interaction involving 
fast collisions of NLS solitons, and demonstrates that this 
interaction plays a key role in collisions in the presence 
of generic nonlinear loss. Moreover, the scalings of $n$-pulse 
interaction effects with $n$ and $m$ and the strong dependence 
on initial positions lead to complex collision dynamics. 
This dynamics is very different from the one encountered in 
fast $N$-soliton collisions in the presence of weak cubic loss, 
where the total collision-induced amplitude shift is a sum over 
amplitude shifts due to two-pulse interaction \cite{PNC2010}.

The rest of the paper is organized as follows. 
In Sec. \ref{Dynamics}, we obtain the reduced model for 
amplitude dynamics in a fast $N$-soliton collision in 
the presence of weak nonlinear loss. We then employ the model 
to calculate the total collision-induced amplitude shift and 
the contribution from $n$-soliton interaction. 
In Sec. \ref{Simu}, we analyze in detail the predictions 
of the reduced model for the amplitude shifts in four-soliton 
and six-soliton collisions. In addition, we compare the analytic 
predictions with results of numerical simulations with the 
perturbed NLS equation. In Sec. \ref{Conclusions}, we present 
our conclusions. Appendix A is devoted to the derivation of the 
equation for the collision-induced change in the soliton's envelope 
due to $n$-pulse interaction in a fast $N$-soliton collision.

\section{Amplitude dynamics in $N$-soliton collisions}
\label{Dynamics}
Consider propagation of soliton pulses of the cubic NLS 
equation in the presence of generic weak nonlinear loss
$L(\psi)$, where $\psi$ is the physical field. In the context 
of propagation of light through optical waveguides, for example,  
$\psi$ is proportional to the envelope of the electric field. 
Assume that $L(\psi)$ can be approximated by $G(|\psi|^{2})\psi$,  
where $G$ is a Taylor polynomial of degree $m_{c}$. 
Thus, we can write: 
\begin{eqnarray}
L(\psi)\simeq G(|\psi|^{2})\psi = 
\sum_{m=0}^{m_{c}}\epsilon_{2m+1}|\psi|^{2m}\psi,
\label{loss}
\end{eqnarray}
where $0\le\epsilon_{2m+1}\ll 1$ for $m\ge 0$.  
We refer to the $m$th summand on the right hand side of 
Eq. (\ref{loss}) as $(2m+1)$-order loss and remark that 
this term is often associated with $(m+1)$-photon absorption \cite{Boyd2008}. 
Under the aforementioned assumption on the loss, the dynamics of 
the pulses is governed by:
\begin{eqnarray}
i\partial_z\psi+\partial_t^2\psi+2|\psi|^2\psi=
-i\sum_{m=0}^{m_{c}}\epsilon_{2m+1}|\psi|^{2m}\psi.
\label{single1}
\end{eqnarray}
Here we adopt the notation used in nonlinear optics, 
in which $z$ is propagation distance and $t$ is time. 
The fundamental soliton solution of the unperturbed 
NLS equation with central frequency $\beta_{j}$ is  
\begin{eqnarray}
\psi_{j}(t,z)\!=\!
\eta_{j}\frac{\exp(i\chi_{j})}{\cosh(x_{j})},
\label{single2}
\end{eqnarray}
where $x_{j}=\eta_{j}\left(t-y_{j}-2\beta_{j} z\right)$, 
$\chi_{j}=\alpha_{j}+\beta_{j}(t-y_{j})+
\left(\eta_{j}^2-\beta_{j}^{2}\right)z$, 
and $\eta_{j}$, $y_{j}$, and $\alpha_{j}$ 
are the soliton amplitude, position, and phase, respectively.

The effects of the nonlinear loss on single pulse propagation 
can be calculated by employing the standard adiabatic perturbation 
theory \cite{Hasegawa95}. This perturbative calculation yields 
the following expression for the rate of change of the soliton 
amplitude  
\begin{eqnarray}&&
\frac{d\eta_{j}(z)}{dz}=-\sum_{m=0}^{m_{c}}\epsilon_{2m+1}
a_{2m+1}\eta_{j}^{2m+1}(z), 
\label{single3}
\end{eqnarray} 
where $a_{2m+1}=(2^{m+1}m!)/[(2m+1)!!]$. The $z$ dependence of the 
soliton amplitude is obtained by integration 
of Eq. (\ref{single3}).

Let us discuss the calculation of the effects of weak nonlinear 
loss on a fast collision between $N$ NLS solitons. The solitons 
are identified by the index $j$, where $1\le j\le N$. 
Since we deal with a fast collision, $|\beta_{j}-\beta_{k}|\gg 1$ 
for any $j \ne k$. The only other assumption of our calculation 
is that $0\le\epsilon_{2m+1}\ll 1$ for $m\ge 0$. 
Under these assumptions, we can employ 
a generalization of the perturbation technique, 
developed in Ref. \cite{PCG2003}, and successfully applied for 
studying fast two-soliton and three-soliton collisions in 
different setups \cite{PCG2003,SP2004,CP2005,CP2008,PNC2010,NP2010,
PC2012,PC2012B}. Note that the generalized technique in the current paper 
is more complicated than the one used in Refs. 
\cite{PCG2003,SP2004,CP2005,CP2008,PNC2010,NP2010,PC2012,PC2012B}.
We therefore provide a brief outline of the main steps in the 
generalized calculation. (1) We first consider  
the effects of $(2m+1)$-order loss, and calculate  
the contribution of $n$-soliton interaction 
with $n\le m+1$ to the collision-induced amplitude shift, 
for a given $n$-soliton combination \cite{max_n}. 
(2) We then add the contributions coming from all 
possible $n$-soliton combinations. This sum is the total 
contribution of $n$-pulse interaction to the amplitude shift in a 
fast collision in the presence of $(2m+1)$-order loss. 
(3) Summing the amplitude shifts calculated in (2) over all 
relevant $m$ values, $1\le m \le m_{c}$, we obtain 
the total contribution of $n$-pulse interaction to the 
amplitude shift in a collision in the presence of {\it generic 
nonlinear loss}. (4) The total collision-induced amplitude shift 
is obtained by summing the amplitude shifts in (3) 
over all  possible $n$-values, $2\le n \le m+1$.

Following this procedure, we first calculate the 
collision-induced change in the amplitude of the 
$j$th soliton due to $(2m+1)$-order loss. 
The dynamics is determined by the following 
perturbed NLS equation 
\begin{eqnarray}
i\partial_z\psi+\partial_t^2\psi+2|\psi|^2\psi=
-i\epsilon_{2m+1}|\psi|^{2m}\psi.
\label{n_body1}
\end{eqnarray} 
We start by considering the amplitude shift of the 
$j$th soliton due to $n$-pulse interaction with solitons 
with indexes $l_{1}, l_{2}, \dots, l_{n-1}$, 
where $1\le l_{j'} \le N$ and $l_{j'} \ne j$ for $1\le j' \le n-1$. 
Employing a generalization of the perturbation method developed 
in Ref. \cite{PCG2003}, we look for an $n$-pulse solution 
of Eq. (\ref{n_body1}) in the form $\psi_{n}=\psi_{j}+\phi_{j}+
\sum_{j'=1}^{n-1}[\psi_{l_{j'}}+\phi_{l_{j'}}]+\dots$, 
where $\psi_{k}$ is the $k$th single-soliton  
solution of Eq. (\ref{n_body1}) with $0<\epsilon_{2m+1}\ll 1$, 
$\phi_{k}$ describes collision-induced effects for the $k$th soliton, 
and the ellipsis represents higher-order terms. 
We then substitute $\psi_{n}$  along with  
$\psi_{j}(t,z)=\Psi_{j}(x_{j})\exp(i\chi_{j})$,   
$\phi_{j}(t,z)=\Phi_{j}(x_{j})\exp(i\chi_{j})$, 
$\psi_{l_{j'}}(t,z)=\Psi_{l_{j'}}(x_{l_{j'}})\exp(i\chi_{l_{j'}})$, 
and $\phi_{l_{j'}}(t,z)=\Phi_{l_{j'}}(x_{l_{j'}})\exp(i\chi_{l_{j'}})$ 
for $j'=1,\dots, n-1$, into Eq. (\ref{n_body1}).  
Since the frequency difference for each soliton pair is large, 
we can employ the resonant approximation, and neglect terms with rapid 
oscillations with respect to $z$. Under this approximation, 
Eq. (\ref{n_body1}) decomposes  into a system of equations for 
the evolution of $\Phi_{j}$ and the $\Phi_{l_{j'}}$.  
[See, for example, Refs. \cite{PNC2010,PC2012}, for a discussion 
of the  cases $n=2$ and $n=3$ for $m=1$ and $m=2$]. 
The system of equations is solved by expanding $\Phi_{j}$  
and each of the $\Phi_{l_{j'}}$ in a perturbation series 
with respect to $\epsilon_{2m+1}$ and $1/|\beta_{l_{j'}}-\beta_{j}|$. 
We focus attention on $\Phi_{j}$ and comment that 
the equations for the $\Phi_{l_{j'}}$ are obtained in a 
similar manner. The only collision-induced effect 
in order $1/|\beta_{l_{j'}}-\beta_{j}|$ is a phase shift  
$\Delta\alpha_{j}=4\sum_{j'=1}^{n-1}\eta_{l_{j'}}/
|\beta_{l_{j'}}-\beta_{j}|$, which  
already exists in the unperturbed collision \cite{Zakharov72}. 
Thus, we find that the main effect of $(2m+1)$-order loss 
on the collision is of order $\epsilon_{2m+1}/|\beta_{l_{j'}}-\beta_{j}|$. 
We denote the corresponding term in the expansion of $\Phi_{j}$  
by $\Phi_{j2}^{(1m)}$, where the first subscript  
stands for the soliton index, the second subscript indicates the
combined order with respect to both $\epsilon_{2m+1}$ and 
$1/|\beta_{l_{j'}}-\beta_{j}|$, and the superscripts represent 
the order in $\epsilon_{2m+1}$ and the order of the nonlinear loss, 
respectively. Furthermore, the contribution to $\Phi_{j2}^{(1m)}$ 
from $n$-soliton interaction with the $l_{1}, l_{2}, \dots, l_{n-1}$  
solitons is denoted by $\Phi_{j2(l_{1}\dots l_{n-1})}^{(1mn)}$. 
In Appendix A, we show that the latter contribution satisfies:  
\begin{eqnarray}&&
\partial_{z}\Phi_{j2(l_{1}\dots l_{n-1})}^{(1mn)}=
-\epsilon_{2m+1}\sum_{k_{l_{1}}=1}^{m-(n-2)}
\sum_{k_{l_{2}}=1}^{m-k_{l_{1}}-(n-3)}\dots
\sum_{k_{l_{n-1}}=1}^{m-s_{n-2}}
\frac{m!(m+1)!}{(k_{l_{1}}! \dots k_{l_{n-1}}!)^{2}}
\nonumber \\&&
\times
\left[(m+1-s_{n-1})!(m-s_{n-1})!\right]^{-1}
|\Psi_{l_{1}}|^{2k_{l_{1}}} \dots |\Psi_{l_{n-1}}|^{2k_{l_{n-1}}}
|\Psi_{j}|^{2m-2s_{n-1}}\Psi_{j},
\label{n_body2}
\end{eqnarray}
where $s_{n}=\sum_{j'=1}^{n}k_{l_{j'}}$.  
Note that all terms in the sum on the right hand side of 
Eq. (\ref{n_body2}) contain the products   
$|\Psi_{l_{1}}|^{2k_{l_{1}}} \dots |\Psi_{l_{n-1}}|^{2k_{l_{n-1}}}
|\Psi_{j}|^{2k_{j}}\Psi_{j}$, where $k_{l_{1}}+ \dots +k_{l_{n-1}}+k_{j}=m$, 
and $1\le k_{l_{j'}} \le m-(n-2)$ for $1\le j'\le n-1$. 
Therefore, the largest value of $n$ that can induce non-vanishing 
effects is obtained by setting $k_{j}=0$ and $k_{l_{j'}}=1$ 
for $1\le j'\le n-1$. This yields $n_{\mbox{max}}=m+1$ for 
the maximum value of $n$.

Next, we obtain the equation for the rate of change of the $j$th  
soliton's amplitude due to $n$-pulse interaction with the 
$l_{1}, l_{2}, \dots, l_{n-1}$ solitons. For this purpose, 
we first expand both sides of Eq. (\ref{n_body2}) with respect to 
the eigenmodes of the linear operator $\hat L$ describing small perturbations 
about the fundamental NLS soliton \cite{PCG2003,SP2004,CP2005,PNC2010,PC2012}. 
We then project the two expansions onto the eigenmode 
$f_{0}(x_{j})=\mbox{sech}(x_{j})(1,-1)^{T}$ and integrate 
over $x_{j}$. This calculation yields the following equation 
for the rate of change of the amplitude: 
\begin{eqnarray}&&
\frac{d\eta_{j(l_{1}\dots l_{n-1})}^{(mn)}}{dz}=
-\epsilon_{2m+1}\sum_{k_{l_{1}}=1}^{m-(n-2)}
\dots\sum_{k_{l_{n-1}}=1}^{m-s_{n-2}}
\frac{m!(m+1)!\eta_{l_{1}}^{2k_{l_{1}}} \dots \eta_{l_{n-1}}^{2k_{l_{n-1}}} 
\eta_{j}^{2m-2s_{n-1}+1}}
{(k_{l_{1}}! \dots k_{l_{n-1}}!)^{2}(m+1-s_{n-1})!(m-s_{n-1})!}
\nonumber \\&&
\times
\int_{-\infty}^{\infty}dx_{j}
[\cosh(x_{l_{1}})]^{-2k_{l_{1}}} \dots [\cosh(x_{l_{n-1}})]^{-2k_{l_{n-1}}}
[\cosh(x_{j})]^{-(2m-2s_{n-1}+2)}.
\label{n_body3}
\end{eqnarray}

We now proceed to the second calculation step,  
in which we obtain the total rate of change in the $j$th soliton's 
amplitude due to $n$-pulse interaction in a fast $N$-soliton 
collision in the presence of $(2m+1)$-order loss. For this purpose, 
we sum Eq. (\ref{n_body3}) over all $n$-soliton combinations 
$(j, l_{1}, \dots, l_{n-1})$, where $1\le l_{j'} \le N$, 
$l_{j'} \ne j$, and $1\le j' \le n-1$. Thus, the total rate 
of change of the amplitude due to $n$-pulse interaction is: 
\begin{eqnarray}&&
\frac{d\eta_{j}^{(nm)}}{dz}=\sum_{l_{1}=1}^{N}
\sum_{l_{2}=l_{1}+1}^{N}
\dots\sum_{l_{n-1}=l_{n-2}+1}^{N}
\Pi_{j'=1}^{n-1}\left(1-\delta_{jl_{j'}}\right)
\frac{d\eta_{j(l_{1}\dots l_{n-1})}^{(mn)}}{dz}, 
\label{n_body5}
\end{eqnarray} 
where $\delta_{jk}$ is the Kronecker delta function.       
The total rate of change in the $j$th soliton's amplitude 
in an $N$-soliton collision in the presence of {\it the generic 
nonlinear loss} due to $n$-soliton interaction 
is calculated by summing both sides of Eq. (\ref{n_body5}) over 
$m$ for $n-1\le m \le m_{c}$. This yields:   
\begin{eqnarray}&&
\frac{d\eta_{j}^{(n)}}{dz}=\sum_{m=n-1}^{m_{c}}
\frac{d\eta_{j}^{(mn)}}{dz}. 
\label{n_body5a}
\end{eqnarray} 
To obtain the total rate of change of the amplitude in the 
collision, we sum Eq. (\ref{n_body5a}) over $n$ for $2\le n \le m_{c}+1$,    
and also take into account the effects of single-pulse propagation, 
as described by Eq. (\ref{single3}). We arrive at the following equation:  
\begin{eqnarray}&&
\frac{d\eta_{j}}{dz}=\sum_{n=2}^{m_{c}+1}
\frac{d\eta_{j}^{(n)}}{dz}
-\sum_{m=0}^{m_{c}}\epsilon_{2m+1}
a_{2m+1}\eta_{j}^{2m+1}, 
\label{n_body6}
\end{eqnarray}        
for $j=1,\dots, N$. Equations (\ref{n_body3})-(\ref{n_body6}) 
provide a complete description of the collision-induced 
amplitude dynamics under the assumptions of a fast collision 
and weak loss.

Important insight into the effects of $n$-pulse interaction 
on $N$-soliton collisions is obtained by studying full-overlap 
collisions, i.e., collisions in which the envelopes 
of all $N$ solitons overlap at a certain distance $z_{c}$. 
More specifically, we would like to calculate the total 
collision-induced amplitude shift $\Delta\eta_{j}$ in 
these collisions, and compare it with the contributions 
of $n$-pulse interaction to the amplitude shift  
$\Delta\eta_{j}^{(n)}$, for $n=2, \dots, m_{c}+1$. 
For this purpose, we consider first a full-overlap 
$N$-soliton collision in the presence of $(2m+1)$-order loss.   
The rate of change in the $j$th soliton's amplitude 
due to $n$-pulse interaction with solitons with indexes 
$l_{1}, l_{2}, \dots, l_{n-1}$, where 
$1\le l_{j'} \le N$ and $l_{j'} \ne j$ for $1\le j' \le n-1$, 
is given by  Eq. (\ref{n_body3}). In a fast full-overlap 
collision in the presence of weak $(2m+1)$-order loss,   
the main contribution to the amplitude shift comes from 
the close vicinity of the collision point $z_{c}$. 
Therefore, an approximate expression for the contribution 
of $n$-pulse interaction to the amplitude shift can be obtained by 
integrating Eq. (\ref{n_body3}) over $z$ from $-\infty$ to $\infty$, 
while taking the amplitude values on the right hand side 
of the equation as constants \cite{approx}: $\eta_{k}=\eta_{k}(z_{c}^{-})$. 
Employing these steps, we arrive at 
\begin{eqnarray}&&
\Delta\eta_{j(l_{1}\dots l_{n-1})}^{(mn)}
=-\epsilon_{2m+1}\sum_{k_{l_{1}}=1}^{m-(n-2)}
\dots\sum_{k_{l_{n-1}}=1}^{m-s_{n-2}}
\frac{m!(m+1)!\eta_{l_{1}}^{2k_{l_{1}}} \dots \eta_{l_{n-1}}^{2k_{l_{n-1}}} 
\eta_{j}^{2m-2s_{n-1}+1}}
{(k_{l_{1}}! \dots k_{l_{n-1}}!)^{2}(m+1-s_{n-1})!(m-s_{n-1})!}
\nonumber \\&&
\!\!\!\!\times\int_{-\infty}^{\infty}dx_{j}
[\cosh(x_{j})]^{-(2m-2s_{n-1}+2)}
\!\!\int_{-\infty}^{\infty}dz
[\cosh(x_{l_{1}})]^{-2k_{l_{1}}} \dots [\cosh(x_{l_{n-1}})]^{-2k_{l_{n-1}}}.
\label{n_body7}
\end{eqnarray}  
The total contribution of $n$-pulse interaction to the 
amplitude shift in a fast full-overlap $N$-soliton collision 
in the presence of $(2m+1)$-order loss is obtained 
by summing Eq. (\ref{n_body7}) over all $n$-soliton 
combinations $(j, l_{1}, \dots, l_{n-1})$: 
\begin{eqnarray}&&
\Delta\eta_{j}^{(mn)}=
\sum_{l_{1}=1}^{N}
\sum_{l_{2}=l_{1}+1}^{N}
\dots\sum_{l_{n-1}=l_{n-2}+1}^{N}
\Pi_{j'=1}^{n-1}\left(1-\delta_{jl_{j'}}\right)
\Delta\eta_{j(l_{1}\dots l_{n-1})}^{(mn)}.
\label{n_body8}
\end{eqnarray} 
Summation of  Eq. (\ref{n_body8}) over $m$ yields the total 
contribution of $n$-pulse interaction to the amplitude shift 
in a full-overlap collision in the presence of the 
{\it generic nonlinear loss}:   
\begin{eqnarray}&&
\Delta\eta_{j}^{(n)}
=\sum_{m=n-1}^{m_{c}}
\Delta\eta_{j}^{(mn)}.
\label{n_body9}
\end{eqnarray}           
Thus, the approximate expression for the 
total amplitude shift in a fast full-overlap collision is 
\begin{eqnarray}&&
\Delta\eta_{j}=\sum_{n=2}^{m_{c}+1}
\Delta\eta_{j}^{(n)}.
\label{n_body10}
\end{eqnarray}       
Note that since Eqs. (\ref{n_body3})-(\ref{n_body6}) 
and Eqs. (\ref{n_body7})-(\ref{n_body10}) are independent 
of the soliton phases, the total collision-induced amplitude shift 
and the contribution of $n$-soliton interaction are expected to 
be phase-insensitive.

\section{Analysis and simulations of four-soliton 
and six-soliton collisions}
\label{Simu}
Let us demonstrate the implications of Eqs. (\ref{n_body3})-(\ref{n_body6}) 
and Eqs. (\ref{n_body7})-(\ref{n_body10}) on collision-induced 
amplitude dynamics in specific setups. We start by analyzing the effects of 
fast full-overlap $N$-soliton collisions in the presence of 
$(2m+1)$-order loss, where the dynamics is described by Eq. (\ref{n_body1}).  
Since the cases $m=1$ and $m=2$ were already studied by us in 
Refs. \cite{PNC2010,PC2012}, we first focus attention on collisions 
in the presence of septic loss ($m=3$). 
As we show below, the analysis of this case is sufficient for uncovering 
the main scaling properties and the importance of $n$-soliton 
interaction in soliton collisions in the presence of high-order loss. 
For concreteness, we consider four-soliton and six-soliton collisions 
with soliton frequencies, $\beta_{1}=0$, $\beta_{2}=-\Delta\beta$, 
$\beta_{3}=\Delta\beta$, $\beta_{4}=2\Delta\beta$ for $N=4$, 
and $\beta_{1}=0$, $\beta_{2}=-2\Delta\beta$, $\beta_{3}=-\Delta\beta$, 
$\beta_{4}=\Delta\beta$, $\beta_{5}=2\Delta\beta$, 
$\beta_{6}=3\Delta\beta$ for $N=6$.          
Note that this choice corresponds, for example, 
to the one used in optical waveguide links employing 
wavelength-division-multiplexing \cite{MM98}. 
The initial amplitudes and phases are $\eta_{j}(0)=1$ and 
$\alpha_{j}(0)=0$ for $1\le j \le N$, respectively.    
The initial positions are $y_{0}(1)=0$, $y_{2}(0)=20$, 
$y_{3}(0)=-20$, $y_{4}(0)=-40$ for $N=4$, 
and $y_{0}(1)=0$, $y_{2}(0)=40$, $y_{3}(0)=20$, $y_{4}(0)=-20$, 
$y_{5}(0)=-40$, $y_{6}(0)=-60$ for $N=6$. Thus, the solitons are 
well separated before the collision. In addition, the final 
propagation distance $z_{f}$ is assumed to be large enough, 
so that the solitons are well separated after the collision. 
The value of the septic loss coefficient is taken 
as $\epsilon_{7}=0.002$.

\begin{figure}[ptb]
\begin{tabular}{cc}
\epsfxsize=8.0cm  \epsffile{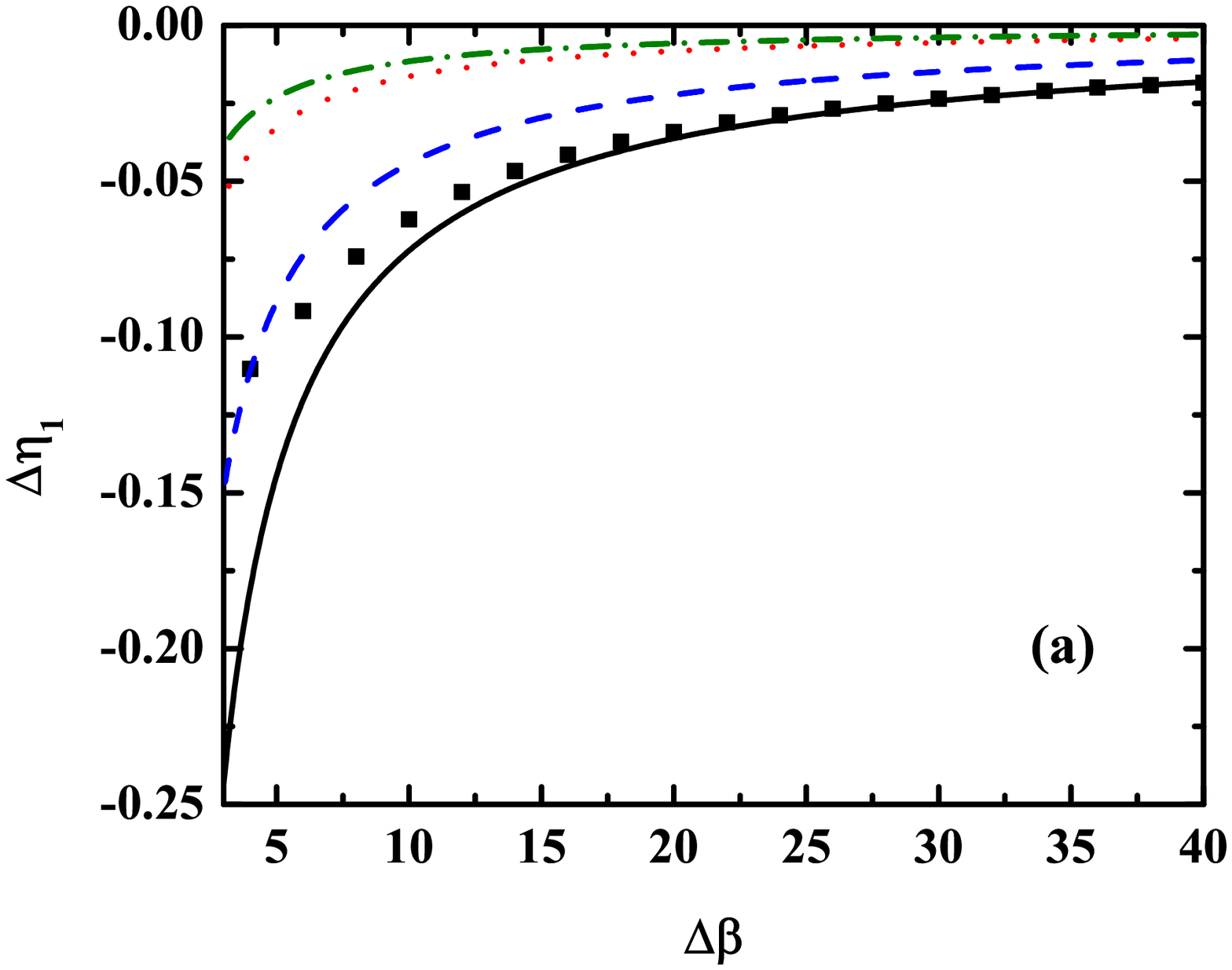} \\
\epsfxsize=8.0cm  \epsffile{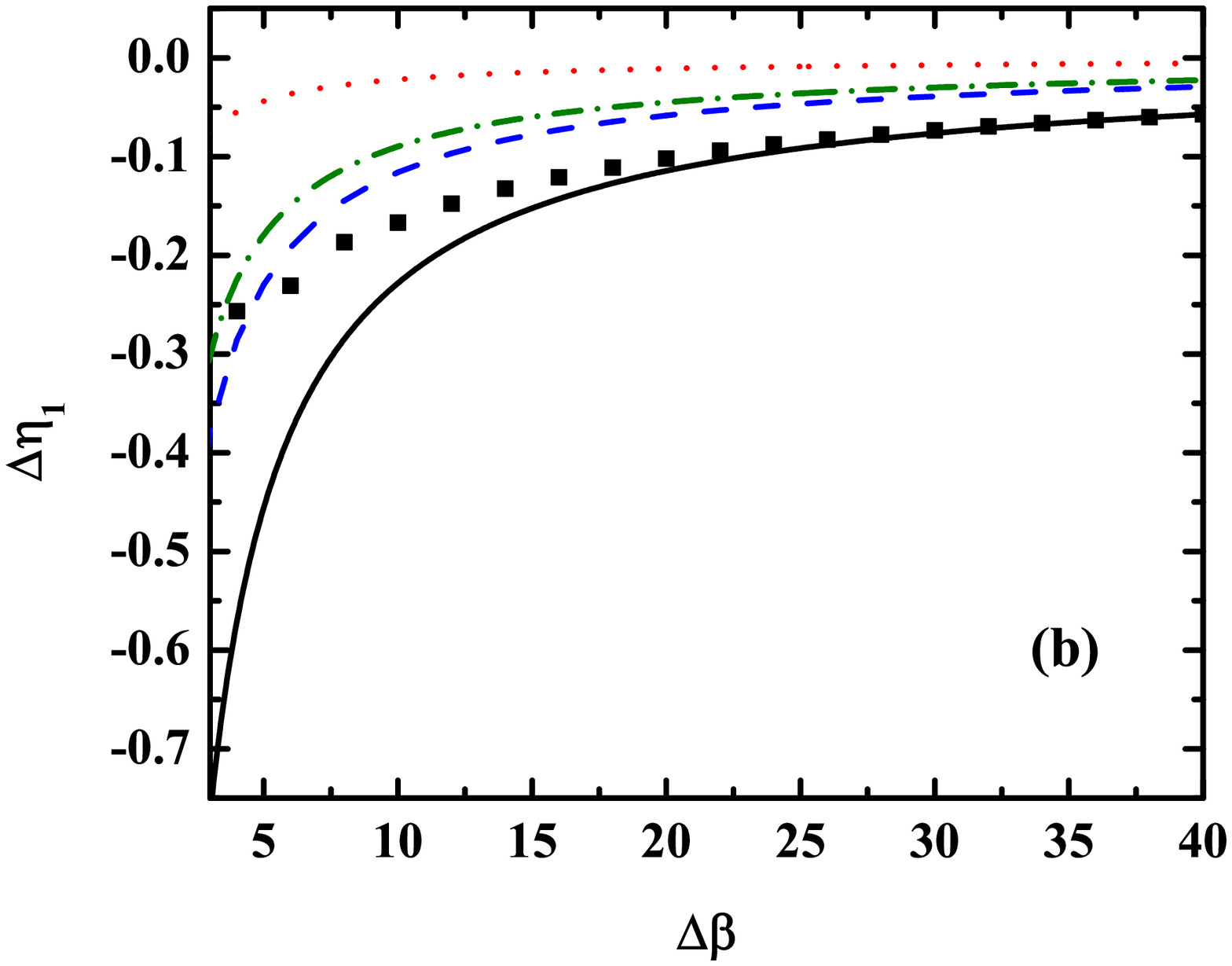}   
\end{tabular}
\caption{(Color online) The total collision-induced amplitude shift 
of the $j=1$ soliton $\Delta\eta_{1}$ vs frequency difference $\Delta\beta$ 
in a full-overlap four-soliton collision (a) 
and in a full-overlap six-soliton collision (b) 
in the presence of septic loss with coefficient $\epsilon_{7}=0.002$. 
The solid black line is the analytic prediction of 
Eqs. (\ref{n_body7})-(\ref{n_body10}) and the squares represent 
the result of numerical simulations with Eq. (\ref{n_body1}). 
The dotted red, dashed blue, and dashed-dotted green lines 
correspond to the contributions of two-, three-, and four-soliton 
interaction to the amplitude shift, $\Delta\eta_{1}^{(2)}$, 
$\Delta\eta_{1}^{(3)}$, and $\Delta\eta_{1}^{(4)}$, respectively.} 
\label{fig1}
\end{figure}

Figure \ref{fig1} shows the $\Delta\beta$-dependence of the total 
collision-induced amplitude shift in four-pulse and six-pulse 
collisions, for the $j=1$ ($\beta_{j}=0$) soliton. 
Both the prediction of Eqs. (\ref{n_body7})-(\ref{n_body10})
and the result obtained by numerical solution of Eq. (\ref{n_body1}) 
are presented. The figure also shows the analytic prediction for 
the contributions of two-, three-, and four-soliton interaction 
to the amplitude shift, $\Delta\eta_{1}^{(2)}$, $\Delta\eta_{1}^{(3)}$, 
and $\Delta\eta_{1}^{(4)}$, respectively.   
The agreement between the analytic prediction and the numerical 
simulations is very good for $\Delta\beta \ge 15$, where the 
perturbation description is expected to hold. Moreover, 
our calculations show that the dominant contribution to the 
total amplitude shift in a four-soliton collision  
comes from three-soliton interaction. 
The contribution from four-soliton interaction increases from 
$15.9\%$ in a four-soliton collision to $39.4\%$ in a 
six-soliton collision. Consequently, in a six-soliton collision 
the effects of three-pulse and four-pulse interaction are both  
important, while those of two-pulse interaction are relatively 
small (about $9.6\%$). Further numerical simulations of fast 
full-overlap four-soliton collisions show that the total 
collision-induced amplitude shift is insensitive to the initial 
phases of the solitons, in agreement with the analytic prediction 
of Eqs. (\ref{n_body7})-(\ref{n_body10}). 
Based on these observations we conclude that 
phase-insensitive $n$-pulse interaction with high $n$ values,  
satisfying $2< n \le m+1$, plays a crucial role in fast 
full-overlap $N$-soliton collisions in the presence 
of $(2m+1)$-order loss.

\begin{figure}[ptb]
\begin{tabular}{cc}
\epsfxsize=8.0cm  \epsffile{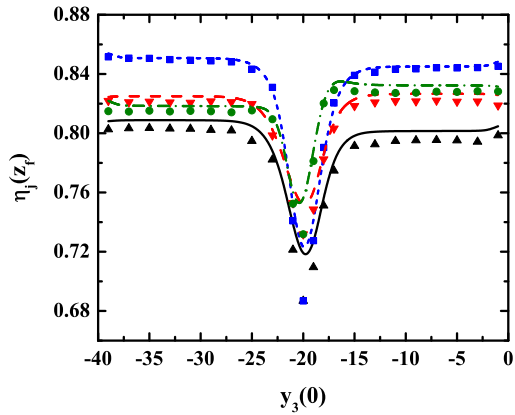} 
\end{tabular}
\caption{(Color online) The final soliton amplitudes $\eta_{j}(z_{f})$ 
vs the initial position of the $j=3$ soliton $y_{3}(0)$ in a 
four-soliton collision in the presence of septic loss 
with $\epsilon_{7}=0.02$. The solid black curve, dashed red curve, 
short-dashed blue curve, and dashed-dotted green curve 
represent the analytic predictions  
of Eqs. (\ref{n_body3})-(\ref{n_body6}) for 
$\eta_{j}(z_{f})$ with $j=1,2,3,4$, respectively. 
The black up triangles, red down triangles, blue squares, 
and green circles correspond to the results obtained by numerical 
solution of Eq. (\ref{n_body1}) for $\eta_{j}(z_{f})$ 
with $j=1,2,3,4$, respectively.} 
\label{fig2}
\end{figure}

We now turn to analyze more generic fast $N$-soliton collisions, 
in which the solitons' envelopes do not completely overlap. 
Based on Eq.  (\ref{n_body3}), the contribution of $n$-pulse 
interaction to the total amplitude shift should strongly 
depend on the degree of soliton overlap during the collision, 
for $n\ge 3$, $m\ge 2$, and $N\ge 3$. 
Consequently, the total collision-induced 
amplitude shift might strongly depend on the initial soliton 
positions in this case. We therefore focus our attention 
on this dependence. We consider, as an example, a four-soliton 
collision in the presence of septic loss with $\epsilon_{7}=0.02$, 
where the soliton frequencies are  $\beta_{1}=0$, $\beta_{2}=-10$, 
$\beta_{3}=10$, and $\beta_{4}=20$.       
The initial amplitudes and phases are $\eta_{j}(0)=1$ and 
$\alpha_{j}(0)=0$ for $1\le j \le 4$. The initial positions 
are $y_{0}(0)=0$, $y_{2}(0)=20$, $y_{4}(0)=-40$, and 
$-39\le y_{3}(0) \le -1$. That is, the initial position of 
the $j=3$ soliton is varied, while the initial positions 
of the other solitons are not changed. 
Notice that in this setup, the four-soliton collision is 
{\it not} a full-overlap collision, except for at $y_{3}(0)=-20$. 
As a result, Eqs.  (\ref{n_body7})-(\ref{n_body10}) 
cannot be employed to analyze the collision-induced amplitude 
dynamics, and Eqs.  (\ref{n_body3})-(\ref{n_body6}) 
should be used instead. We therefore solve 
Eqs. (\ref{n_body3})-(\ref{n_body6}) with the aforementioned  
initial parameter values for $0\le z \le z_{f}$, where $z_{f}=6$, 
and plot the final amplitudes $\eta_{j}(z_{f})$ vs $y_{3}(0)$. 
The curves are shown in Fig. (\ref{fig2}) along with the 
curves obtained by numerical solution of Eq. (\ref{n_body1}). 
The agreement between the analytic prediction and the 
simulations result is good. As can be seen, each 
$\eta_{j}(z_{f})$-vs-$y_{3}(0)$ curve has a pronounced 
minimum at $y_{3}(0)=-20$, i.e., at the initial position value  
of the $j=3$ soliton corresponding to a full-overlap 
collision. Thus, a strong dependence of the collision-induced 
amplitude shift on the initial soliton positions is observed 
already in a four-soliton collision in the presence of 
septic loss. This means that the collision-induced amplitude 
dynamics in fast $N$-soliton collisions in the presence of weak 
generic loss can be quite complex due to the dominance of contributions 
from $n$-pulse interaction with high $n$-values. This behavior 
is sharply different from the one encountered in fast $N$-soliton 
collisions in the presence of weak cubic loss. In the latter 
case, the total collision-induced amplitude shift is a sum over contributions 
from two-pulse interaction, and the collision can be accurately viewed 
as consisting of a collection of pointwise two-soliton 
collisions \cite{PNC2010}.

\begin{figure}[ptb]
\begin{tabular}{cc}
\epsfxsize=8.0cm  \epsffile{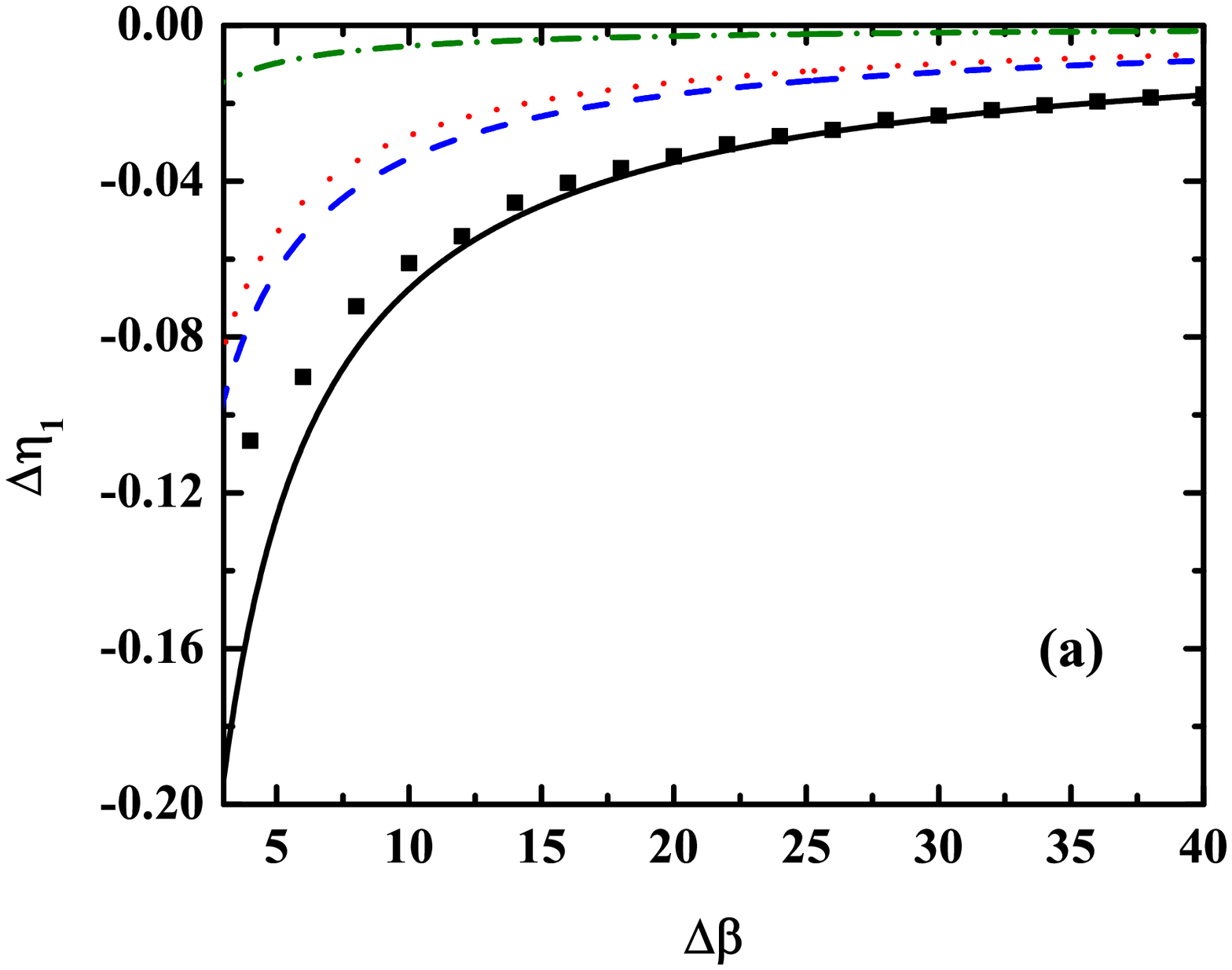} \\
\epsfxsize=8.0cm  \epsffile{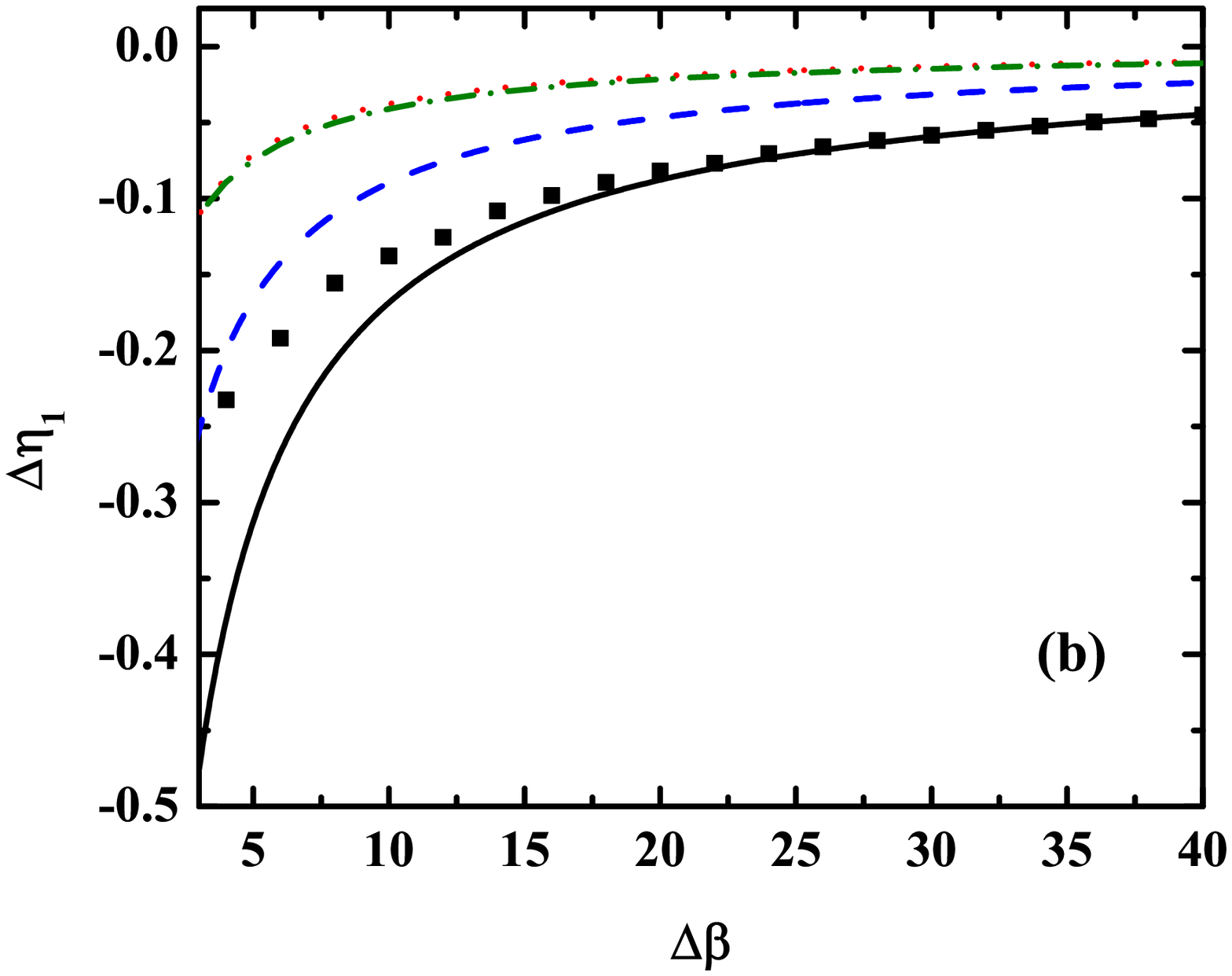}   
\end{tabular}
\caption{(Color online) The total collision-induced amplitude shift 
of the $j=1$ soliton $\Delta\eta_{1}$ vs frequency difference $\Delta\beta$ 
in a full-overlap four-soliton collision (a) 
and in a full-overlap six-soliton collision (b) 
in the presence of generic nonlinear loss of the form 
(\ref{loss}) with $m_{c}=3$ and loss coefficients 
$\epsilon_{1}=0.002$, $\epsilon_{3}=0.004$, 
$\epsilon_{5}=0.006$, and $\epsilon_{7}=0.001$.   
The solid black line is the analytic prediction of 
Eqs. (\ref{n_body7})-(\ref{n_body10}) and the squares correspond to 
the result of numerical simulations with Eq. (\ref{single1}). 
The dotted red, dashed blue, and dashed-dotted green lines 
represent the contributions of two-, three-, and four-soliton 
interaction to the amplitude shift, $\Delta\eta_{1}^{(2)}$, 
$\Delta\eta_{1}^{(3)}$, and $\Delta\eta_{1}^{(4)}$, respectively.} 
\label{fig3}
\end{figure}

The analysis of the effects of $(2m+1)$-order loss on 
$N$-soliton collisions is very valuable, since it explains 
the importance of $n$-pulse interaction and uncovers the 
scaling laws for this interaction. However, in most systems 
one has to take into account the impact of the low-order 
loss terms, whose presence can enhance the effects of 
two-pulse interaction. It is therefore important to 
take into account all the relevant loss terms when analyzing 
collision-induced dynamics in the presence of generic loss. 
We now turn to address this aspect of the problem, 
by considering the effects of generic weak nonlinear loss 
of the form (\ref{loss}) on fast $N$-soliton collisions.   
For concreteness, we assume $m_{c}=3$ and 
loss coefficients $\epsilon_{1}=0.002$, $\epsilon_{3}=0.004$, 
$\epsilon_{5}=0.006$, and $\epsilon_{7}=0.001$. 
We also assume full-overlap collisions, but emphasize that 
the analysis can be extended to treat the general case by 
the same method described in the preceding paragraph. 
We consider four-soliton and six-soliton collisions 
with the same pulse parameters used for full-overlap collisions 
in the presence of septic loss. Figure \ref{fig3} shows 
the $\Delta\beta$ dependence of the total collision-induced amplitude 
shift in four-soliton and six-soliton collisions for the $j=1$ soliton, 
as obtained by Eqs. (\ref{n_body7})-(\ref{n_body10}).
The result obtained by numerical solution of Eq. (\ref{single1})  
and the analytic predictions for the contributions of 
two-, three-, and four-soliton interaction, 
$\Delta\eta_{1}^{(2)}$, $\Delta\eta_{1}^{(3)}$, 
and $\Delta\eta_{1}^{(4)}$, are also shown.       
We observe that in four-soliton collisions, 
$\Delta\eta_{1}^{(2)}$ is comparable to $\Delta\eta_{1}^{(3)}$, 
while $\Delta\eta_{1}^{(4)}$ is much smaller. 
That is, the inclusion of the low-order loss terms does lead 
to an enhancement of the fractional contribution of two-pulse  
interaction to the amplitude shift. In contrast, 
in six-soliton collisions, $\Delta\eta_{1}^{(3)}$ ($53.2\%$) is 
significantly larger than $\Delta\eta_{1}^{(2)}$ ($22.2\%$), 
while $\Delta\eta_{1}^{(4)}$ ($24.6\%$) is comparable 
to $\Delta\eta_{1}^{(2)}$. 
Based on the latter observation we conclude that when the 
low-order loss coefficients $\epsilon_{1}$ and 
$\epsilon_{3}$ are comparable in magnitude to the higher-order 
loss coefficients, the contributions to the amplitude shift from 
$n$-pulse interaction with $n\ge 3$ can be much larger than 
the one coming from two-pulse interaction.

\section{Conclusions}
\label{Conclusions}
In summary, we studied $n$-pulse interaction in fast collisions 
of $N$ solitons of the cubic NLS equation in the presence of generic 
weak nonlinear loss, which can be approximated by the series (\ref{loss}). 
Due to the presence of nonlinear loss, the solitons 
experience collision-induced amplitude shifts that are strongly 
enhanced by $n$-pulse interaction. We first developed a general 
reduced model that allowed us to calculate the contribution of 
$n$-pulse interaction to the amplitude shift in fast $N$-soliton 
collisions in the presence of $(2m+1)$-order loss, for any $n$ 
and $m$. We then used the reduced model and numerical simulations 
with the perturbed NLS equation to analyze four-soliton and 
six-soliton collisions in the presence of septic loss $(m=3)$. 
Our calculations showed that three-pulse interaction gives the 
dominant contribution to the collision-induced amplitude shift 
already in a full-overlap four-soliton collision, while in a full-overlap 
six-soliton collision, both three-pulse and four-pulse interaction 
are important. Furthermore, we found that the collision-induced 
amplitude shift has a strong dependence on the initial soliton positions, 
with a pronounced maximum in the case of a full-overlap collision. 
We then generalized these results by considering $N$-soliton 
collisions in the presence of generic weak nonlinear loss of the 
form (\ref{loss}) with $m_{c}=3$. Our analytic calculations and numerical 
simulations showed that three-pulse interaction gives the 
dominant contribution to the amplitude shift in a full-overlap 
six-soliton collision, despite the presence of linear and 
cubic loss. All the collision-induced effects were found to be 
insensitive to the soliton phases for fast collisions. 
Based on these observations we conclude that phase-insensitive 
$n$-pulse interaction with high $n$ values plays a key role 
in fast collisions of NLS solitons in the presence of generic weak  
nonlinear loss. The complex scalings of $n$-pulse interaction 
effects with $n$ and $m$ and the strong dependence on initial 
soliton positions lead to complex collision dynamics. 
This dynamics is very different from the one observed in 
fast collisions of $N$ NLS solitons in the presence of weak cubic loss, 
where the total collision-induced amplitude shift is a sum over 
amplitude shifts due to two-pulse interaction \cite{PNC2010}. 
          
\section*{Acknowledgments}
Q.M. Nguyen is supported by the Vietnam National Foundation for Science 
and Technology Development (NAFOSTED) under grant number 101.02-2012.10.

\appendix
\section{Derivation of Eq. (\ref{n_body2})}
In this Appendix, we derive Eq. (\ref{n_body2}) for    
the collision-induced change in the envelope of a soliton 
due to $n$-pulse interaction in a fast $N$-soliton 
collision in the presence of weak $(2m+1)$-order loss. 
More specifically, we consider the change in the envelope 
of the $j$th soliton induced by $n$-pulse interaction with 
solitons with indexes $l_{1}, l_{2}, \dots, l_{n-1}$, 
where $1\le l_{j'} \le N$ and $l_{j'} \ne j$ for $1\le j' \le n-1$. 
The derivation is based on a generalization of the perturbation 
procedure developed in Ref. \cite{PCG2003}. Following this procedure, 
we look for a solution of Eq. (\ref{n_body1}) in the form 
$\psi_{n}=\psi_{j}+\phi_{j}+
\sum_{j'=1}^{n-1}[\psi_{l_{j'}}+\phi_{l_{j'}}]+\dots$, 
where $\psi_{k}$ is the $k$th single-soliton  
solution of Eq. (\ref{n_body1})  with $0<\epsilon_{2m+1}\ll 1$, 
$\phi_{k}$ describes collision-induced effects for the $k$th soliton, 
and the ellipsis represents higher-order terms. 
We then substitute $\psi_{n}$  along with  
$\psi_{j}(t,z)=\Psi_{j}(x_{j})\exp(i\chi_{j})$,   
$\phi_{j}(t,z)=\Phi_{j}(x_{j})\exp(i\chi_{j})$, 
$\psi_{l_{j'}}(t,z)=\Psi_{l_{j'}}(x_{l_{j'}})\exp(i\chi_{l_{j'}})$, 
and $\phi_{l_{j'}}(t,z)=\Phi_{l_{j'}}(x_{l_{j'}})\exp(i\chi_{l_{j'}})$ 
for $j'=1,\dots, n-1$, into Eq. (\ref{n_body1}).      
Next, we use the resonant approximation, and neglect terms with rapid 
oscillations with respect to $z$. We find that the main effect of 
$(2m+1)$-order loss on the envelope of the $j$th soliton is of order 
$\epsilon_{2m+1}/|\beta_{l_{j'}}-\beta_{j}|$.  
We denote this collision-induced change in the envelope by 
$\Phi_{j2}^{(1m)}$, and the contribution to this change from 
$n$-soliton interaction with the $l_{1}, l_{2}, \dots, l_{n-1}$  
solitons by $\Phi_{j2(l_{1}\dots l_{n-1})}^{(1mn)}$. 
Within the resonant approximation, the phase factor of 
terms contributing  to changes in the $j$th soliton's envelope 
must be equal to $\chi_{j}$. Consequently, these terms must be 
proportional to: $|\Psi_{l_{1}}|^{2k_{l_{1}}} \dots 
|\Psi_{l_{n-1}}|^{2k_{l_{n-1}}}|\Psi_{j}|^{2k_{j}}\Psi_{j}$, 
where $k_{l_{1}}+ \dots +k_{l_{n-1}}+k_{j}=m$, and 
$1\le k_{l_{j'}} \le m-(n-2)$ for $1\le j'\le n-1$.
Summing over all possible contributions of this form, 
we obtain the following evolution equation for 
$\Phi_{j2(l_{1}\dots l_{n-1})}^{(1mn)}$: 
\begin{eqnarray}&&
\partial_{z}\Phi_{j2(l_{1}\dots l_{n-1})}^{(1mn)}=
-\epsilon_{2m+1}\sum_{k_{l_{1}}=1}^{m-(n-2)}
\sum_{k_{l_{2}}=1}^{m-k_{l_{1}}-(n-3)}\dots
\sum_{k_{l_{n-1}}=1}^{m-s_{n-2}}
b_{\mathbf{k}}
\nonumber \\&&
\times
|\Psi_{l_{1}}|^{2k_{l_{1}}} \dots |\Psi_{l_{n-1}}|^{2k_{l_{n-1}}}
|\Psi_{j}|^{2m-2s_{n-1}}\Psi_{j}, 
\label{append1}
\end{eqnarray}    
where $s_{n}=\sum_{j'=1}^{n}k_{l_{j'}}$, 
$b_{\mathbf{k}}$ are constants, and 
$\mathbf{k}=(k_{l_{1}},k_{l_{2}},\dots,k_{l_{n-1}})$.

To calculate the expansion coefficients $b_{\mathbf{k}}$, 
we first note that 
\begin{eqnarray}&&
|\Psi|^{2m}\Psi=
\left(\Psi_{j}+\sum_{j'=1}^{n-1}\Psi_{l_{j'}}\right)^{m+1} 
\left(\Psi_{j}^{*}+\sum_{j'=1}^{n-1}\Psi_{l_{j'}}^{*}\right)^{m}.
\label{append2}
\end{eqnarray}  
Employing the multinomial expansion formula for the two terms 
on the right hand side of Eq. (\ref{append2}), we obtain: 
\begin{eqnarray}&&
\left(\Psi_{j}+\sum_{j'=1}^{n-1}\Psi_{l_{j'}}\right)^{m+1}= 
\sum_{k_{l_{1}}=0}^{m+1}\dots\sum_{k_{l_{n-1}}=0}^{m+1} 
\frac{(m+1)!}{(k_{l_{1}}! \dots k_{l_{n-1}}!)(m+1-s_{n-1})!}
\nonumber \\&&
\times
\Psi_{l_{1}}^{k_{l_{1}}} \dots \Psi_{l_{n-1}}^{k_{l_{n-1}}}
\Psi_{j}^{m+1-s_{n-1}},
\label{append3}
\end{eqnarray}   
and 
\begin{eqnarray}&&
\left(\Psi_{j}^{*}+\sum_{j'=1}^{n-1}\Psi_{l_{j'}}^{*}\right)^{m}= 
\sum_{k_{l_{1}}=0}^{m}\dots\sum_{k_{l_{n-1}}=0}^{m} 
\frac{m!}{(k_{l_{1}}! \dots k_{l_{n-1}}!)(m-s_{n-1})!}
\nonumber \\&&
\times
\Psi_{l_{1}}^{*k_{l_{1}}} \dots \Psi_{l_{n-1}}^{*k_{l_{n-1}}}
\Psi_{j}^{*m-s_{n-1}}.
\label{append4}
\end{eqnarray}   
Combining Eqs. (\ref{append2})-(\ref{append4}), we find that 
the expansion coefficients $b_{\mathbf{k}}$ are given by: 
\begin{eqnarray}&&
b_{\mathbf{k}}=
\frac{m!(m+1)!}{(k_{l_{1}}! \dots k_{l_{n-1}}!)^{2}
(m+1-s_{n-1})!(m-s_{n-1})!}. 
\label{append5}
\end{eqnarray}  
Substituting this relation into Eq. (\ref{append2}), we arrive 
at Eq. (\ref{n_body2}).


\begin{thebibliography}{}

\bibitem{Roy2005} A.E. Roy, 
{\it Orbital Motion} (Institute of Physics, Bristol, 2005). 

\bibitem{Hagihara70} Y. Hagihara, {\it Celestial Mechanics I: 
Dynamical Principles and Transformation Theory}  
(MIT Press, Cambridge, MA, 1970). 

\bibitem{Thouless72} D.J. Thouless, 
{\it The Quantum Mechanics of Many-Body Systems}  
(Academic, New York, 1972). 

\bibitem{Definition} Here we use the term $n$-body interaction 
(or force) to describe an interaction (or a force) that does not 
exist in a system with $k$ objects, where $2\le k \le n-1$, 
but does appear in an $n$-body system.

\bibitem{Teller43} B.M. Axilrod and E. Teller, 
J. Chem. Phys. {\bf 11}, 299 (1943).

\bibitem{Fujita57} J. Fujita and H. Miyazawa, Prog. Theor. Phys. {\bf 17}, 
360 (1957). 

\bibitem{Nogami67} B.A. Loiseau and Y. Nogami, Nucl. Phys. B {\bf 2}, 
470 (1967).

\bibitem{Witala98} H. Witala, W. Gl\"ockle, D. H\"uber, J. Golak, 
and H. Kamada, Phys. Rev. Lett. {\bf 81}, 1183 (1998). 

\bibitem{Witala2002} E. Epelbaum, A. Nogga, W. Gl\"ockle, 
H. Kamada, U.G. Meissner, and H. Witala, 
Phys. Rev. C {\bf 66}, 064001 (2002).

\bibitem{Daley2009} A.J. Daley, J.M. Taylor, S. Diehl, M. Baranov, 
and P. Zoller, Phys. Rev. Lett. {\bf 102}, 040402 (2009). 

\bibitem{Honer2010} J. Honer, H. Weimer, T. Pfau, and H.P. B\"uchler, 
Phys. Rev. Lett. {\bf 105}, 160404 (2010). 

\bibitem{Liang2012} Y. Liang and H. Guo, J. Phys. B {\bf 45}, 
175303 (2012). 


\bibitem{Zakharov84} S. Novikov, S.V. Manakov, L.P. Pitaevskii, 
and V.E. Zakharov, {\it Theory of Solitons: The Inverse Scattering Method}  
(Plenum, New York, 1984). 

\bibitem{Newell85} A.C. Newell, {\it Solitons in Mathematics and Physics} 
(SIAM, Philadelphia, 1985).

\bibitem{Dalfovo99} F. Dalfovo, S. Giorgini, L.P. Pitaevskii, 
and S. Stringari, Rev. Mod. Phys. {\bf 71}, 463 (1999).  

\bibitem{BEC2008} R. Carretero-Gonz\'alez, D.J. Frantzeskakis, 
and P.G. Kevrekidis, Nonlinearity {\bf 21}, R139 (2008).  

\bibitem{Hasegawa95} A. Hasegawa and Y. Kodama, 
{\it Solitons in Optical Communications} (Clarendon, Oxford, 1995). 

\bibitem{Agrawal2001} G.P. Agrawal, {\it Nonlinear 
Fiber Optics} (Academic, San Diego, CA, 2001).

\bibitem{Asano69} N. Asano, T. Taniuti, and N. Yajima, 
J. Math. Phys. {\bf 10}, 2020 (1969)

\bibitem{Ichikawa72} Y.H. Ichikawa, T. Imamura, and T. Taniuti, 
J. Phys. Soc. Jpn. {\bf 33}, 189 (1972). 

\bibitem{Horton96} W. Horton and Y.H. Ichikawa, 
{\it Chaos and Structure in Nonlinear Plasmas}  
(World Scientific, Singapore, 1996).    

\bibitem{Malomed89} Y.S. Kivshar and B.A. Malomed,   
Rev. Mod. Phys. {\bf 61}, 763 (1989). 

\bibitem{Boyd2008} R.W. Boyd, {\it Nonlinear Optics} 
(Academic, San Diego, CA, 2008). 

\bibitem{Burt97} E.A. Burt, R.W. Ghrist, C.J. Myatt, M.J. Holland, 
E.A. Cornell, and C.E. Wieman, Phys. Rev. Lett. {\bf 79}, 337 (1997). 
 
\bibitem{PK2007} K.M. Mertes, J.W. Merrill, R. Carretero-Gonz\'alez, 
D.J. Frantzeskakis, P.G. Kevrekidis, and D.S. Hall, 
Phys. Rev. Lett. {\bf 99}, 190402 (2007).

\bibitem{Kramer2002} I.S. Aranson and L. Kramer, Rev. Mod. Phys.
{\bf 74}, 99 (2002).


\bibitem{PNC2010} A. Peleg, Q.M. Nguyen, and Y. Chung, 
Phys. Rev. A {\bf 82}, 053830 (2010).   

\bibitem{PC2012} A. Peleg and Y. Chung, 
Phys. Rev. A {\bf 85}, 063828 (2012).    

\bibitem{PCG2003} A. Peleg, M. Chertkov, and I. Gabitov, 
Phys. Rev. E {\bf 68}, 026605 (2003).

\bibitem{SP2004} J. Soneson and A. Peleg, Physica D {\bf 195}, 123 (2004). 

\bibitem{CP2005} Y. Chung and A. Peleg, Nonlinearity {\bf 18}, 1555 (2005). 

\bibitem{CP2008} Y. Chung and A. Peleg, Phys. Rev. A  {\bf 77}, 
063835 (2008). 

\bibitem{NP2010} Q.M. Nguyen and A. Peleg,  
J. Opt. Soc. Am. B {\bf 27}, 1985 (2010).  

\bibitem{PC2012B} A. Peleg and Y. Chung, Opt. Commun. {\bf 285}, 
1429 (2012).  


\bibitem{max_n} Note that terms with higher $n$ values do 
not contribute due to fast oscillations with respect to $z$. 
See, for example, \cite{PNC2010}.


\bibitem{Zakharov72} V.E. Zakharov and A.B. Shabat, 
Zh. Eksp. Teor. Fiz. {\bf 61}, 118 (1971) 
[Sov. Phys. JETP {\bf 34}, 62 (1972)].

\bibitem{approx} See Refs. \cite{PNC2010,PC2012}, 
for an application of this approximation for the cases 
$m=1$ and $m=2$, respectively.     

\bibitem{MM98} L.F. Mollenauer and P.V. Mamyshev,
IEEE J. Quantum Electron. {\bf 34}, 2089 (1998). 

\end{thebibliography}
\end{document}